# Hydex Glass: a CMOS Compatible Platform for Integrated Waveguide Structures for Nonlinear Optics

David Duchesne[1], Marcello Ferrera[1], Luca Razzari[1],
Roberto Morandotti[1], Brent Little[2], Sai T. Chu[2] & David J. Moss[3]

[1]*INRS-EMT,*
*Canada*
[2]*Infinera Corporation,*
*USA*
[3]*IPOS/CUDOS, School of Physics, University of Sydney,*
*Australia*

## 1. Introduction

Integrated photonic technologies are rapidly becoming an important and fundamental milestone for wideband optical telecommunications. Future optical networks have several critical requirements, including low energy consumption, high efficiency, greater bandwidth and flexibility, which must be addressed in a compact form factor [1-3]. In particular, it has become well accepted that devices must possess a CMOS compatible fabrication procedure in order to exploit the large existing silicon technology in electronics [4, 5]. This would primarily serve to reduce costs by developing hybrid electro-optic technologies on-chip for ultrafast signal processing. There is still however, a growing demand to implement all-optical technologies on these chips for frequency conversion [6, 7], all-optical regeneration [8,9], multiplexing and demultiplexing [10-12], as well as for routing and switching [10,12]. The motivation for optical technologies is primarily based on the ultrahigh bandwidth of the optical fiber and the extremely low attenuation coefficient. Coupled with minimal pulse distortion properties, such as dispersion and nonlinearities, optical fibers are the ideal transmission medium to carry information over long distances and to connect optical networks. Unfortunately, the adherence of the standard optical fiber to pulse distortions is also what renders it less than perfectly suited for most signal processing applications required in telecommunications. Bending losses become extremely high in fibers for chip-scale size devices, limiting its integrability in networks. Moreover, its weak nonlinearity limits the practical realization (i.e. low power values and short propagation lengths) of some fundamental operations requiring nonlinear optical phenomena, such as frequency conversion schemes and switching [13].

Several alternative material platforms have been developed for photonic integrated circuits [1,2,14,15], including semiconductors such as AlGaAs and silicon-on-insulator (SOI) [3,5,14,16], as well as nonlinear glasses such as chalcogenides, silicon oxynitride and bismuth oxides [1,17,18]. In addition, exotic and novel manufacturing processes have led to new and promising structures in these materials and in regular silica fibers. Photonic crystal fibers [19], 3D photonic bandgap structures [20], and nanowires [21] make use of the tight light confinement to enhance nonlinearities, greatly reduce bending radii, which allows for submillimeter photonic chips.

Despite the abundance of alternative fabrication technologies and materials, there is no clear victor for future all-optical nonlinear devices. Indeed, many nonlinear platforms require power levels that largely exceed the requirements for feasible applications, whereas others have negative side effects such as saturation and multi-photon absorption. Moreover, there is still a fabrication challenge to reduce linear attenuation and to achieve CMOS compatibility for many of these tentative photonic platforms and devices. In response to these demands, a new high-index doped silica glass platform was developed in 2003 [22], which combines the best of both the qualities of single mode fibers, namely low propagation losses and robust fabrication technology, and those of semiconductor materials, such as the small quasi-lossless bending radii and the high nonlinearity. This book chapter primarily describes this new material platform, through the characterization of its linear and nonlinear properties, and shows its application for all-optical frequency conversion for future photonic integrated circuits. In section 2 we present an overview of concurrent recent alternative material platforms and photonic structures, discussing advantages and limitations. We then review in section 3 the fundamental equations for nonlinear optical interactions, followed by an experimental characterization of the linear and nonlinear properties of a novel high-index glass. In section 4 we introduce resonant structures and make use of them to obtain a highly efficient all-optical frequency converter by means of pumping continuous wave light.

## 2. Material platforms and photonic structures for nonlinear effects

### 2.1 Semiconductors

Optical telecommunications is rendered possible by carrying information through waveguiding structures, where a higher index core material ($n_c$) is surrounded by a cladding region of lower index material ($n_s$). Nonlinear effects, where the polarization of media depends nonlinearly on the applied electric field, are generally observed in waveguides as the optical power is increased. Important information about the nonlinear properties of a waveguide can be obtained from the knowledge of the index contrast ($\Delta n = n_c - n_s$) and the index of the core material, $n_c$. The strength of nonlinear optical interactions is predominantly determined through the magnitude of the material nonlinear optical susceptibilities ($\chi^{(2)}$ and $\chi^{(3)}$ for second order and third order nonlinear processes where the permittivity depends on the square and the cube of the applied electromagnetic field, respectively), and scales with the inverse of the effective area of the supported waveguide mode. Through Miller's rule (Boyd, 2008) the nonlinear susceptibilities can be shown to depend almost uniquely on the refractive index of the material, whereas the index contrast can easily be used to estimate the area of the waveguide mode, where a large index contrast leads to a more confined (and thus a smaller area) mode. It thus comes to no surprise that the most commonly investigated materials for nonlinear effects are III-V semiconductors, such as silicon and AlGaAs, which possess a large index of refraction at the telecommunications wavelength ($\lambda$ = 1.55 μm) and where waveguides with a large index contrast can be formed. For third order nonlinear phenomena such as the Kerr effect[1], the strength of the nonlinear interactions can be estimated through the nonlinear parameter $\gamma = n_2\omega/cA$ [13], where $n_2$ is the nonlinear index coefficient determined solely from material properties, $\omega$ is the angular frequency of the light, $c$ is the speed of light and $A$ the effective area of the mode, which will be more clearly defined later. The total cumulative nonlinear effects induced by a waveguide sample can be roughly estimated as being proportional to the peak power, length of the waveguide and the nonlinear parameter [13]. In order to minimize the energetic requirements, it is thus necessary either to have long structures and/or large nonlinear parameters. Focusing on the moment on the nonlinear parameter, in typical semiconductors, the core index $n_c$ > 3 (~3.5 for Si and ~3.3 GaAs) leads to values of $n_2$ ~$10^{-18}$ – $10^{-17}$ m$^2$/W, to be compared with fused silica ($n_c$ = 1.45) where $n_2$ ~2.6 x $10^{-20}$ m$^2$/W. Moreover, etching through the waveguide core allows for a large index contrast with air, permitting photonic wire geometries with effective areas below 1 um$^2$, see Fig. 1. This leads to extremely high values of $\gamma$ ~ 200,000W$^{-1}$km$^{-1}$ [8,21] (to be compared with single mode fibers which have $\gamma$ ~ 1W$^{-1}$km$^{-1}$ [13]). This large nonlinearity has been used to demonstrate several nonlinear applications for telecommunications, including all-optical regeneration at 10 Gb/s using four-wave mixing and self-phase modulation in SOI [8,23], frequency conversion [6,7,24], and Raman amplifications [25, 26].

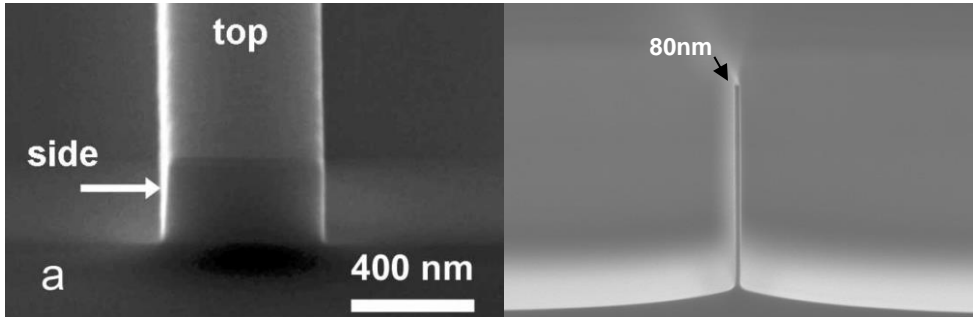

Fig. 1. (left) Silicon-on-insulator nano-waveguide (taken from [21]) and inverted nano-taper (80nm in width) of an AlGaAs waveguide (right). Both images show the very advanced fabrication processes of semiconductors.

There are however major limitations that still prevent their implementation in future optical networks. Semiconductor materials typically have a high material dispersion (a result of being near the bandgap of the structure), which prevents the fabrication of long structures. To overcome this problem, small nano-size wire structures, where the waveguide dispersion dominates, allows one to tailor the total induced dispersion. The very advanced fabrication technology for both Si and AlGaAs allows for this type of control, thus a precise waveguide geometry can be fabricated to have near zero dispersion in the spectral regions of interest.

---

[1] We will neglect second order nonlinear phenomena, which are not possible in centrosymmetric media such as glasses. See [50] and [7, 71] for recent advances in exploiting $\chi^{(2)}$ media for optical telecommunications.

Unfortunately, the small size of the mode also implies a relatively large field along the waveguide etched sidewalls (see Fig. 1). This leads to unwanted scattering centers and surface state absorptions where initial losses have been higher than 10dB/cm for AlGaAs [27-29], and ~ 3 dB/cm for SOI [6].

Another limitation comes from multiphoton absorption (displayed pictorially in Fig. 2 for the simplest case, i.e. two-photon absorption) and involves the successive absorption of photons (via virtual states) that promotes an electron from the semiconductor valence band to the conduction band. This leads to a saturation of the transmitted power and, consequently, of the nonlinear effects. For SOI this has been especially true, where losses are not only due to two-photon absorption, but also to the free carriers induced by the process [21, 30]. Moreover, the nonlinear figure of merit (= $n_2/\alpha_2\lambda$, where $\alpha_2$ is the two photon absorption coefficient), which determines the feasibility of nonlinear interactions and switching, is particular low in silicon [31].

Lastly, although reducing the modal area enhances the nonlinear properties of the waveguide, it also impedes coupling from the single mode fiber into the device; for comparison the modal diameter of a fiber is ~10μm whereas for a nanowire structure it is typically 20 times smaller. This leads to high insertion losses through the device, necessitating either expensive amplifiers at the output, or of complicated tapers often requiring mature fabrication technologies and sometimes multi-step etching processes [32] (SOI waveguides make use of state-of-the-art inverse tapers which limits the insertion losses to approximately 5dB [6, 33].

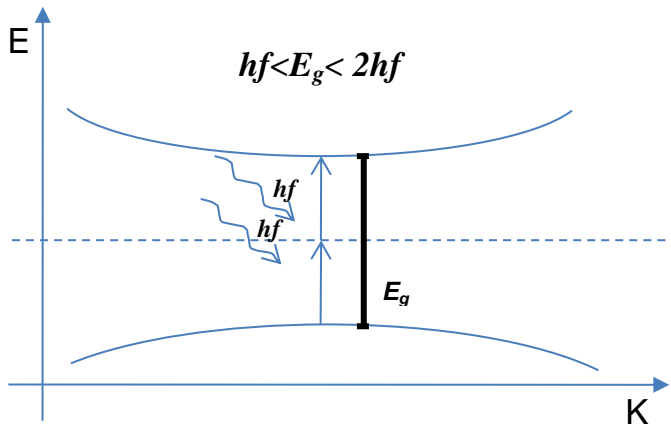

Fig. 2. Schematic of two-photon absorption in semiconductors. In the most general case of the multiphoton absorption process, electrons pass from the valence band to the conduction band via the successive absorption of multiple photons, mediated via virtual states, such that the total absorbed energy surpasses the bandgap energy.

## 2.2 High index glasses

In addition to semiconductors, a number of high index glass systems have been investigated as a platform for future photonic integrated networks, including chalcogenides [1, 17], silicon nitride [34] and silicon oxynitride [35]. Chalcogenides in particular have been shown to have extremely high nonlinear parameters approaching $\gamma \sim 100{,}000 W^{-1} km^{-1}$ in nanotapers [36], which has been used to demonstrate demultiplexing at 160 Gb/s [37]. However, all of these platforms suffer from shortcomings of one form or another. Fabrication processes for chalcogenide glasses are still under development [38, 39] and while they generally possess a very high nonlinear figure of merit - significantly better than silicon, for example - it can be an issue for some glasses [40]. Photosensitivity and photo-darkening, while powerful tools for creating novel photonic structures, can sometimes place limits on the material stability [41]. Whereas other high-index glasses, such as silicon oxynitride, have negligible nonlinear absorption (virtually infinite figure of merit), high temperature annealing is required to reduce propagation losses, making the entire process non-CMOS compatible.

A high-index, doped silica glass material called Hydex® [22], was developed by Little Optics in 2003 as a compromise between the attractive linear features of silica glass and the nonlinear properties of semiconductors. Films are first deposited using standard chemical vapour deposition. Subsequently, waveguides are formed using photolithography and reactive ion etching, producing waveguide sidewalls with exceptionally low roughness. The waveguides are then buried in standard fused silica glass, making the entire fabrication process CMOS compatible and requiring no further anneal. The typical waveguide cross section is 1.45 x 1.5 μm² as shown in Fig. 3. The linear index at $\lambda$ = 1.55 μm is 1.7, and propagation losses have been shown to be as low as 0.06 dB/cm [42, 43]. In addition, fiber pigtails have been designed for coupling to and from Hydex waveguides, with coupling losses on the order of 1.5dB. The linear properties of this material platform has already been exploited to achieve filters with >80dB extinction ratios [44], as well as the optical sensing of biomolecules [45].

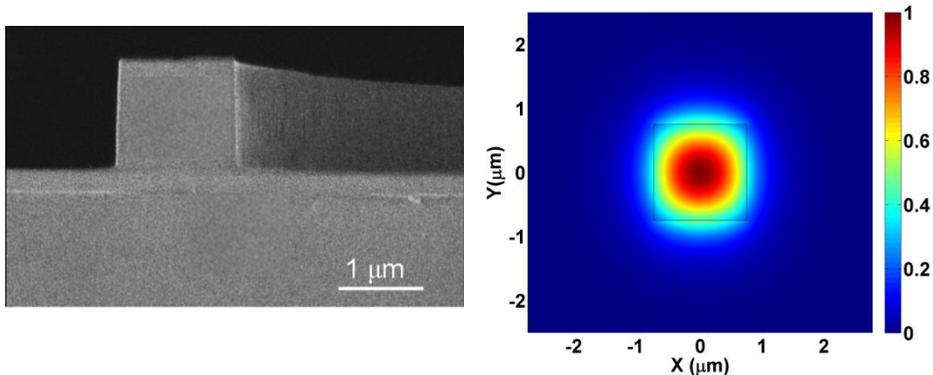

Fig. 3. Scanning electron microscopy picture of the high-index glass waveguide (prior to upper $SiO_2$ deposition), and electromagnetic field distribution of the fundamental mode.

As will be shown in the subsequent sections below, this material platform also has a moderate nonlinearity, and coupled with long or resonant structures, can be used to generate significant nonlinear effects with low power requirements. In the next section we produce the necessary equations governing light propagation in a nonlinear media, followed by a characterization method for the nonlinearity, and explain the possible applications achievable by exploiting resonant and long structures.

## 3. Light dynamics in nonlinear media

In order to completely characterize the nonlinear optical properties of materials, it is worthwhile to review some fundamental equations relating to pulse propagation in nonlinear media. In general, this is modelled directly from Maxwell's equations, and for piecewise homogenous media one can arrive at the optical nonlinear Schrodinger equation [13, 46]:

$$\frac{\partial \psi}{\partial z} + \beta_1 \frac{\partial \psi}{\partial t} + i\frac{\beta_2}{2}\frac{\partial^2 \psi}{\partial t^2} + HOD + \frac{\alpha_1}{2}\psi = i\gamma|\psi|^2\psi - \frac{\alpha_2}{2A}|\psi|^2\psi - HOL \qquad (1)$$

Where $\psi$ is the slowly-varying envelope of the electric field, given by:

$E = \psi'(z,t)F(x,y)\exp(i\beta_0 z - i\omega_0 t)$, where $\psi'$ has been normalized such that $|\psi|^2$ represents the optical power. $\omega_0$ is the central angular frequency of the pulse, $\beta_0$ the propagation constant, $\beta_1$ is the inverse of the group velocity, $\beta_2$ the group velocity dispersion, $\alpha_1$ the linear loss coefficient, $\alpha_2$ the two-photon absorption coefficient, $\gamma$ (= $n_2\omega_0/cA$) the nonlinear parameter, $t$ is time and $z$ is the propagation direction. Here $F(x,y)$ is the modal electric field profile, which can be found by solving the dispersion relation:

$$\nabla^2 F + \frac{\omega^2 n^2}{c^2}F = \beta^2 F \qquad (2)$$

The eigenvalue solution to the dispersion relation can be obtained by numerical methods such as vectorial finite element method (e.g. Comsol Multiphysics). From this the dispersion parameters can be calculated via a Taylor expansion:

$$\beta = \beta_0 + \beta_1(\omega - \omega_0) + \frac{\beta_2}{2}(\omega - \omega_0)^2 + \frac{\beta_3}{6}(\omega - \omega_0)^3 + ... \qquad (3)$$

The effective area can also be evaluated:

$$A = \frac{\left[\iint_\infty |F|^2 dxdy\right]^2}{\iint_\infty |F|^4 dxdy} \qquad (4)$$

In arriving to eq. (1), we neglected higher order nonlinear contributions, non-instantaneous responses (Raman) and non-phase matched terms; we also assumed an isotropic cubic medium, as is the case for glasses. These approximations are valid for moderate power values and pulse durations down to ~100fs for a pulse centered at 1.55 μm [13]. The terms *HOL* and *HOD* refer to higher order losses and higher order dispersion terms, which may be important in certain circumstances [21, 27]. Whereas eq. (1) also works as a first order model for semiconductors, a more general and exact formulation can be found in [46]. Given the material dispersion properties (found either experimentally or from a Sellmeier model [47]), the only unknown parameters in Eq. (1) are the nonlinear parameter $\gamma$ (or $n_2$ to be more precise), the linear propagation loss coefficient $\alpha_1$ and the nonlinear loss term $\alpha_2$.

The solution to the nonlinear Schrodinger equation has been studied in detail [13, 48]. Here we present the solution to this equation for doped silica glass at low and high power regimes. This allows a complete characterization of the waveguide properties which will be extremely useful in studying nonlinear applications such as frequency conversion.

As will be shown below, one of the several advantages of high-index doped silica glass is in its mature fabrication technology which allows for long waveguides with minimal losses. As can be readily seen in Fig. 4, long spiral waveguides of more than 1m of length can be contained in a 2.5 x 2.5 mm$^2$ area.

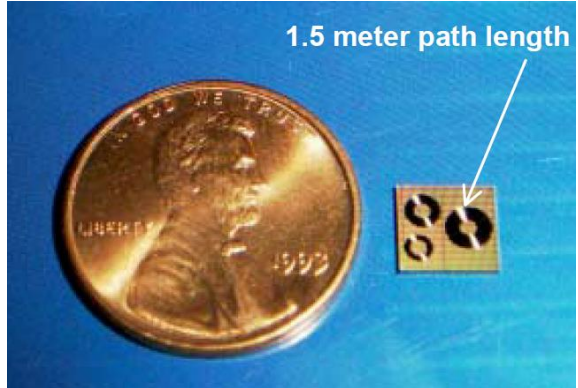

Fig. 4. A 1.5 meter long waveguide confined on a photonic chip smaller than the size of a penny.

### 3.1 Low power regime

At low power, dispersive terms dominate thus leading to temporal pulse broadening. In this limit, the nonlinear Schrodinger equation reduces to:

$$\frac{\partial \psi}{\partial z} + \beta_1 \frac{\partial \psi}{\partial t} + i\frac{\beta_2}{2}\frac{\partial^2 \psi}{\partial t^2} + HOD + \frac{\alpha_1}{2}\psi = 0 \quad (5)$$

This equation transforms to a simple linear ordinary differential equation in the Fourier domain, and assuming an input unchirped Gaussian pulse of width $T_0$ the solution (neglecting $HOD$ terms) is given by [13]:

$$\psi = \frac{T_0}{\sqrt{T_0^2 - i\beta_2 z}}\exp\left(-\frac{\alpha_1}{2}z\right)\exp\left(-\frac{(t-\beta_1 z)^2}{2(T_0^2 - i\beta_2 z)}\right) \quad (6)$$

The pulse is seen to acquire a chirp, leading to temporal broadening via dispersion. The analogue in the spectral domain is that the pulse acquires a quadratic phase, which can serve as a direct measurement of the dispersion induced from the waveguide. A well known experimental technique for reconstructing the phase and amplitude at the output of a device is the Fourier Transform Spectral Interferometry (FTSI) [49]. Using this spectral interference technique, the dispersion of the 45cm doped silica glass spiral waveguide was determined to be very small (on the order of the single mode fiber dispersion, $\beta_2$~22ps$^2$/km), and not important for pulses as short as 100fs [42]. This is extremely relevant, as 3 critical conditions must be met to allow propagation through long structures (note that waveguides are typically <1cm): 1) low linear propagation loses, so that a useful amount of power remains after propagation; 2) low dispersion value so that picosecond pulses or shorter are not broadened significantly; and 3) long waveguides must be contained in a small chip for integration, as was done in the spiral waveguide discussed. This latter requirement also imposes a minimal index contrast $\Delta n$ on the waveguide, such that bending losses are also minimized. Moreover, as will be discussed further below, having a low dispersion value is critical for low power frequency conversion.

**3.2 Nonlinear losses**

In order to see directly the effects of the nonlinear absorption on the propagation of light pulses, it is useful to transform Eq. (1) to a peak intensity equation, $I = |\psi|^2 / A$, as follows:

$$\frac{dI}{dz} = \frac{\psi^*}{A}\frac{\partial \psi}{\partial z} + \frac{\psi}{A}\frac{\partial \psi^*}{\partial z} = -\alpha_1 I - \alpha_2 I^2 - \sum_n \alpha_n I^n, \qquad (7)$$

where we have neglected dispersion contributions based on the previous considerations. We have also explicitly added the higher order multiphoton contributions (three-photon absorption and higher), although it is important to note that these higher order effects typically have a very small cross section that require large intensity values (see chapter 12 of [50]). Considering only two-photon absorption, the solution is found to be:

$$I = \frac{\alpha I_0 \exp(-\alpha z)}{\alpha + \alpha_2 I_0 (1 - \exp(-\alpha z))} \qquad (8)$$

From this one can immediately conclude that the maximal output intensity is limited by two-photon absorption to be $1/\alpha_2 z$; a similar saturation behaviour is obtained when considering higher order contributions. Multiphoton absorption is thus detrimental for high intensity applications and cannot be avoided by any kind of waveguide geometry [46, 50].

Experimentally, the presence of multiphoton absorption can be understood from simple transmission measurements of high power/intensity pulses. Pulsed light from a 16.9MHz Pritel fiber laser, centered at 1.55µm, was used to characterize the transmission in the doped silica glass waveguides. An erbium doped fiber amplifier was used directly after the laser to achieve high power levels, and the estimated pulse duration was approximately 450fs. Fig. 5 presents a summary of the results, showing a purely linear transmission up to input peak powers of 500W corresponding to an intensity of 25GW/cm$^2$ [42]. This result is extremely impressive, and is well above the threshold for silicon [30, 31, 51], AlGaAs [27], or even Chalcogenides [52]. Multiphoton absorption leads to free carrier generation, which in turn can also dramatically increase the losses [30, 31, 51]. For the case of two-photon absorption, the impact on nonlinear signal processing is reflected in the nonlinear figure of merit, $FOM = n_2 / \lambda \alpha_2$, which estimates the maximal Kerr nonlinear contribution with limitations arising from the saturation of the power from two-photon absorption. In high-index doped silica glass, this value is virtually infinite for any practical intensity values, but can be in fact quite low for certain chalcogenides [52] and even lower in silicon (~0.5) [31].

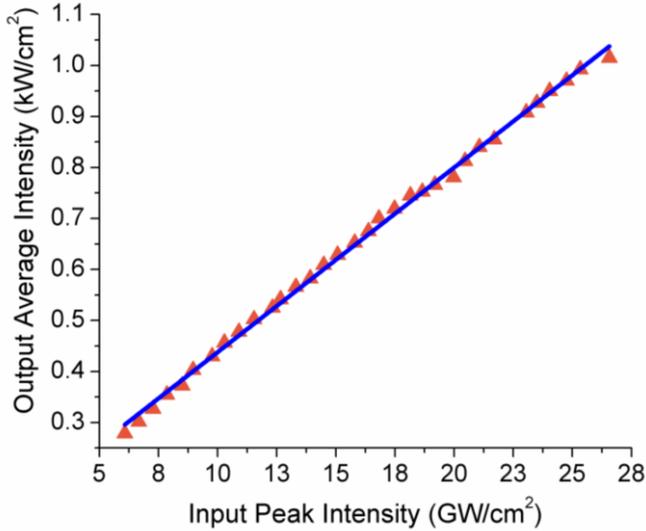

Fig. 5. Transmission at the output of a 45cm long high-index glass waveguide. The linear relation testifies that no multi-photon absorption was present up to peak intensities of more than 25GWcm² (~500W).

By propagating through different length waveguides, we were able to determine, by means of a cut-back style like procedure, both the pigtail losses and propagation losses to be 1.5dB and 0.06dB/cm, respectively. Whereas this value is still far away from propagation losses in single mode fibers (0.2dB/km), it is orders of magnitude better than in typical integrated nanowire structures, where losses >1dB/cm are common [6,27,30]. The low losses, long spiral waveguides confined in small chips, and low loss pigtailing to single mode fibers testifies to the extremely well established and mature fabrication process of this high-index glass platform.

**3.3 Kerr nonlinearity**

In the high power regime, the nonlinear contributions become important in Eq. (1), and in general the equation must be solved numerically. To gain some insight on the effect of the nonlinear contribution to Eq. (1), it is useful to look at the no-dispersion limit of Eq. (1), which can be readily solved to obtain:

$$\psi = \psi_0 \exp\left[i\gamma|\psi_0|^2 \alpha_1^{-1}(1-\exp(-\alpha_1 z))\right] \qquad (9)$$

The nonlinear term introduces a nonlinear chirp in the temporal phase, which in the frequency domain corresponds to spectral broadening (i.e. the generation of new frequencies). This phenomenon, commonly referred to as self-phase modulation, can be used to measure the nonlinear parameter $\gamma$ by means of recording the spectrum of a high power pulse at the output of a waveguide [27, 30, 42]. The nonlinear interactions are found to scale with the product of the nonlinear parameter $\gamma$, the peak power of the pulse, and the effective length of the waveguide (reduced from the actual length due to the linear losses). For low-loss and low-dispersion guiding structures, it is thus useful to have long structures in order to increase the total accumulated nonlinearity, while maintaining low peak power levels. It will be shown in the next section how resonant structures can make use of this to achieve impressive nonlinear effects with 5mW CW power values. For other applications, dispersion effects may be desired, such as for soliton formation [53].

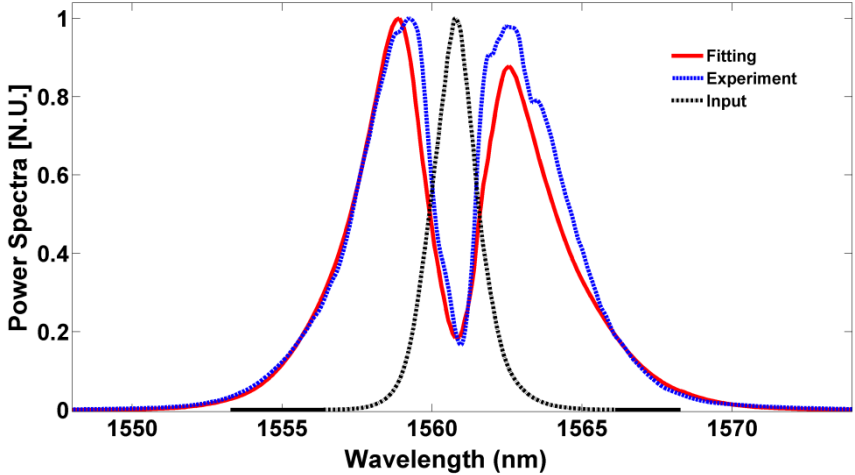

Fig. 6. Input (black) and output spectra (blue) from the 45cm waveguide. Spectral broadening is modelled via numerical solution of Eq. (1) (red curve).

Experimentally, the nonlinearity of the doped silica glass waveguide was characterized in [42] by injecting 1.7ps pulses (centered at 1.55μm) with power levels of approximately 10-60W. The output spectrum showed an increasing amount of spectral broadening, as can be seen in Fig. 6. The value of the nonlinearity was determined by numerically solving the nonlinear Schrodinger equation by means of a split-step algorithm [13], where the only unknown parameter was the nonlinear parameter. By fitting experiments with simulations, a value of $\gamma$ = 220 $W^{-1}km^{-1}$ was determined, corresponding to a value of $n_2$ = 1.1 x $10^{-19}$ $m^2$/W ($A$ = 2.0 μm²). Similar experiments in single mode fibers [13, 54], semiconductors [27, 30], and chalcogenides [52] were also performed to characterize the Kerr nonlinearity. In comparison, the value of $n_2$ obtained in doped silica glass is approximately 5 times larger than that found in standard fused silica, consistent with Miller's rule [50]. On the other hand, the obtained $\gamma$ value is more than 200 times larger, due to the much smaller effective mode area of the doped silica waveguide in contrast to the weakly guided single mode fiber. However, semiconductors and chalcogenides nanotapers definitely have the upper hand in terms of bulk nonlinear parameter values, where $\gamma$ ~ 200,000 $W^{-1}km^{-1}$ have been reported [21,36], due to both the smaller effective mode areas and the larger $n_2$, as previously mentioned.

From Eq. (9), there are 2 ways to improve the nonlinear interactions (for a fixed input power): 1) increasing the nonlinear parameter, or 2) increasing the propagation length. To increase the former, one can reduce the modal size by having high-index contrast waveguides, and/or using a high index material with a high value of $n_2$. Thus, for nonlinear applications, the advantage for doped silica glass waveguides lies in exploiting its low loss and advanced fabrication processes that yield long winding structures, which is typically not possible in other material platforms due to nonlinear absorptions and/or immature fabrication technologies.

## 4. Resonant structures

Advances in fabrication processes and technologies have allowed for the fabrication of small resonant structures whereby specific frequencies of light are found to be "amplified" (or resonate) inside the resonator [55]. Resonators have found a broad range of applications in optics, including high-order filters [44], as oscillators in specific parametric lasers [56, 57], thin film polarization optics, and for frequency conversion [6, 43]. For the case of nonlinear optics, disks (whispering gallery modes) and micro-ring resonators have been used in 2D for frequency conversion [12, 58], whereas micro-toroids and microsphere have been explored in 3D [56, 59]. The net advantage of these structures is that, for resonant frequencies, a low input optical power can lead to enormous nonlinear effects due to the field enhancement provided by the cavity. In this section we examine the specific case of waveguide micro-ring resonators for wavelength conversion via parametric four wave mixing. Micro-ring resonators are integrated structures which can readily be implemented in future photonic integrated circuits. First a brief review of the field enhancement provided by resonators shall be presented, followed by the four-wave mixing relations. Promising experimental results in high-index doped silica resonators will then be shown and compared with other platforms.

### 4.1 Micro-ring resonators

Consider the four port micro-ring resonator portrayed in Fig. 7, and assume continuous wave light is injected from the Input port. Light is coupled from the input (bus) waveguide into the ring structure via evanescent field coupling [60]. As light circulates around the ring structure, there is net loss from propagation losses, loss from coupling from the ring to the bus waveguides (2 locations), and net gain when the input light is coupled from the bus at the input to the ring. Note that this is in direct analogy with a standard Fabry-Perot cavity, where the reflectivity of the mirrors/sidewalls has been replaced with coupling coefficients. Using reciprocity and energy conservation relations at the coupling junction, the total transmission from the Input port to the Drop port is found to be [55]:

$$I_{Drop} = I_0 \frac{k^2 \exp(-\alpha L/2)}{1+(1-k)^2 \exp(-\alpha L)-2(1-k)\exp(-\alpha L/2)\cos(\beta L)} \qquad (10)$$

Where $I_0$ is the input intensity, $L$ the ring circumference ($=2\pi R$), and $k$ is the power coupling ratio from the bus waveguide to the ring structure. A typical transmission profile inside such a resonator is presented in Fig. 8, where we have also defined the free spectral range and the width of the resonance, $\Delta f_{FW}$. Resonance occurs at frequencies $f_{res}=mc/nL$, where $m$ is an even integer, and $n$ is the effective refractive index of the mode, whereas the free spectral range is given by $FSR = c/nL$. In general the ring resonances are not equally spaced with frequency, as dispersion causes a shift in the index of refraction. The coupling coefficient can be expressed in terms of experimentally measured quantities:

$$k = 1 - \left(\rho + 1 - \sqrt{\rho^2 + 2\rho}\right)\exp(\alpha L/2) \qquad (11)$$

where $\rho \approx \frac{1}{2}\left(\pi\frac{\Delta f_{FW}}{FSR}\right)^2$. At resonance, the local intensity inside the resonator is enhanced due to constructive interference. This intensity enhancement factor can be expressed as:

$$IE = \frac{k}{\left[(1-k)\exp(-\alpha L/2) - 1\right]^2} \qquad (12)$$

These equations have extremely important applications. From Eq. (10) the transmission through the resonator is found to be unique at specific frequencies, hence the device can be utilized as a filter. Even more importantly for nonlinear optics, for an input signal that matches a ring resonance, the intensity is found to be enhanced, which can be utilized to observe large nonlinear phenomena with low input power levels [43]. In the approximation of low propagation losses, Eq. (12) results in $IE \approx \pi \cdot Q \cdot FSR/f_0$, which implies that the larger the ring Q-factor ($Q = f_0/\Delta f_{FW}$), the larger the intensity enhancement.

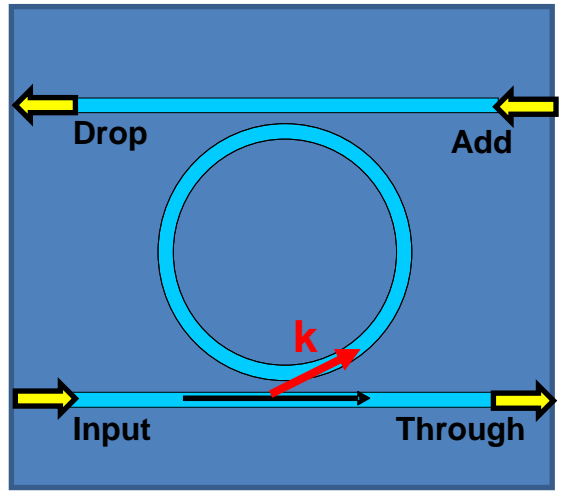

Fig. 7. Coupling coefficients and schematic of a typical 4-port ring resonator.

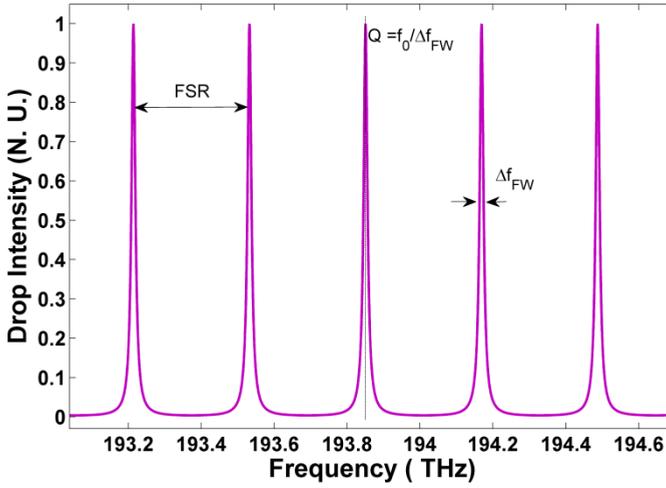

Fig. 8. Typical Fabry-Perot resonance transmission at the drop port of a resonator (input port excited). Here a FSR of 500GHz and a Q of 25 were used.

**4.2 Four-wave mixing**

Section 3 discussed third order nonlinear effects following the propagation of a single beam in a Kerr nonlinear medium. In this case the nonlinear interaction consisted of generating new frequencies through the spectral broadening of the input pulse. In general, we may consider multiple beams propagating through the medium, from which the nonlinear Schrodinger equation predicts nonlinear coupling amongst the components, a parametric process known as four wave mixing. This process can be used to convert energy from a strong pump to generate a new frequency component via the interaction with a weaker signal. As is displayed in the inset of Fig. 9, the quantum description of the process is in the simultaneous absorption of two photons to create 2 new frequencies of light. In the semi-degenerate case considered here, two photons from a strong pump beam ($\psi_2$) are absorbed by the medium, and when stimulated by a weaker signal beam ($\psi_1$) a new idler frequency ($\psi_3$) is generated from the parametric process. By varying the signal frequency, a tunable output source can be obtained [58, 59]. To describe the interaction mathematically, we consider 3 CW beams $E_i = \psi'(z) F(x,y) \exp(i\beta_i z - i\omega_i t)$, from which the following coupled set of equations governing the parametric growth can be derived [13]:

$$\frac{\partial \psi_1}{\partial z} + \frac{\alpha}{2}\psi_1 = 2i\gamma|\psi_2|^2 \psi_1 + i\gamma \psi_2^2 \psi_3^* \exp(i\Delta\beta z) \tag{13a}$$

$$\frac{\partial \psi_2}{\partial z} + \frac{\alpha}{2}\psi_2 = i\gamma|\psi_2|^2 \psi_2 + 2i\gamma \psi_1 \psi_2^* \psi_3 \exp(-i\Delta\beta z) \tag{13b}$$

$$\frac{\partial \psi_3}{\partial z} + \frac{\alpha}{2}\psi_3 = 2i\gamma|\psi_2|^2 \psi_3 + i\gamma \psi_2^2 \psi_1^* \exp(i\Delta\beta z) \tag{13c}$$

Where $\Delta\beta = 2\beta_2 - \beta_3 - \beta_1$ represents the phase mismatch of the process. In arriving to these equations we have assumed that the pump beam ($\omega_2$) is much stronger than the signal ($\omega_1$) and idler ($\omega_3$), and that the waves are closely spaced in frequency so that the nonlinear parameter $\gamma = n_2\omega_0/cA$ is approximately constant for all three frequencies (the pump frequency should be used for $\omega_0$). The phase mismatch term represents a necessary condition (i.e. $\Delta\beta = 0$) for an efficient conversion, and is the optical analogue of momentum conservation. On the other hand, energy conservation is also required and is expressed as: $2\omega_2 = \omega_1 + \omega_3$.

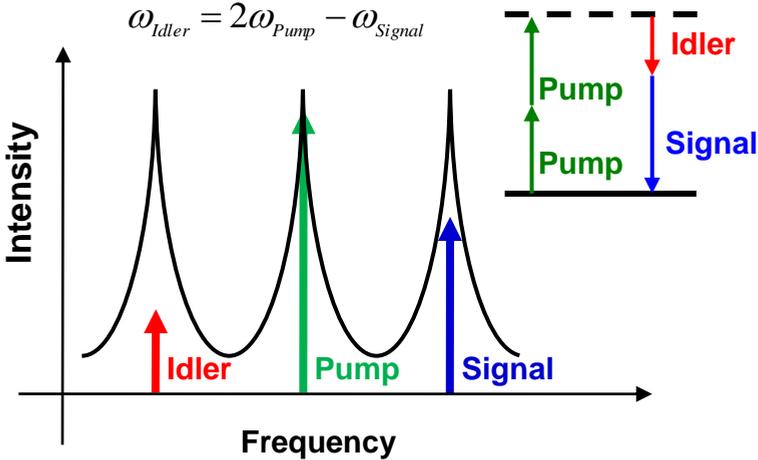

Fig. 9. Typical spectral intensity at the output of the resonator. (Inset) Energy diagram for a semi-degenerate four-wave mixing interaction.

The growth of the idler frequency can be obtained by assuming an un-depleted pump regime, whereby the product $\psi_2 \exp(-\alpha z/2)$ is assumed to be approximately constant, and by solving the Eqs. (13) [13, 24] we obtain:

$$P_3(z) = P_1 \gamma^2 P_2^2 L_{eff}^2 \tag{14a}$$

$$L_{eff}^2 = z^2 \exp(-\alpha z) \left| \frac{1 - \exp(-\alpha z + i\Delta\beta z)}{\alpha z - i\Delta\beta z} \right|^2 \tag{14b}$$

Where $P_1$, $P_2$ and $P_3$ here refer to the input powers of the signal, pump and idler beams respectively. The conversion is seen to be proportional to both the input signal power, and the square of the pump power. Again, we see that the process scales with the nonlinear parameter and is reduced if phase matching (i.e. $\Delta\beta = 0$) is not achieved. Various methods exist to achieve phase matching, including using birefringence and waveguide tailoring [61 - 63], but perhaps the simplest way is to work in a region of low dispersion. As is shown in [13], the phase mismatch term can be reduced to:

$$\Delta\beta \approx \beta_2 (\omega_2 - \omega_1)^2 \tag{15}$$

and is thus directly proportional to the dispersion coefficient (note that at high power levels the phase mismatch becomes power dependant; see [65].

For micro-ring resonator structures, Eq. (14) can be modified to account for the power enhancement provided by the resonance geometry. When the pump and signal beams are aligned to ring resonances, and for low dispersion conditions, phase matching will be obtained, and moreover, the generated idler should also match a ring resonance. In this case we may use Eq. (12) with Eq. (14) to give the expected conversion efficiency:

$$\eta \equiv \frac{P_3(L)}{P_1} = \gamma^2 P_2^2 L_{eff}^2 \cdot IE^4, \qquad (16)$$

where $P_3$ is the power of the idler at the drop port of the ring, whereas $P_1$ and $P_2$ are the input powers both at the Input port, both at the Add port, or one at the Add and the other at the Input (various configurations are possible). The added benefit of a ring resonator for four-wave mixing is clear: the generated idler power at the output of the ring is amplified by a factor of $IE^4$, which can be an extremely important contribution as will be shown below.

Four-wave mixing is an extremely important parametric process to be used in optical networks, and has found numerous applications. This includes the development of a multi-wavelength source for wavelength multiplexing systems [58], all-optical reshaping [65], amplification [62], correlated photon pair generation [66], and possible switching schemes have also been suggested [64]. In particular, signal regeneration using four-wave mixing was shown in silicon at speeds of 10Gb/s [67]. In an appropriate low loss material platform, ring resonators promise to bring efficient parametric processes at low powers.

**4.3 Frequency conversion in doped silica glass resonators**

The possibility of forming resonator structures primarily depends on the developed fabrication processes. In particular, low loss structures are a necessity, as photons will see propagation losses from circulating several times around the resonator. Furthermore, integrated ring resonators require small bending radii with minimal losses, which further require a relatively high-index contrast waveguide. The high-index doped silica glass discussed in this chapter meets these criteria, with propagation losses as low as 0.06 dB/cm, and negligible bending losses for radii down to 30 μm [22, 43].

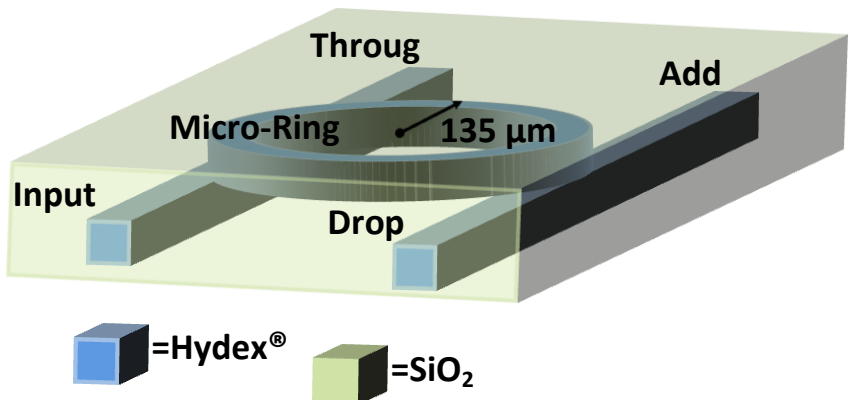

Fig. 10. Schematic of the vertically coupled high-index glass micro-ring resonator.

Two ring resonators will be discussed in this section, one with a radius of 47.5 μm, a Q factor ~65,000, and a bandwidth matching that for 2.5Gb/s signal processing applications, as well as a high Q ring of ~1,200,000, with a ring radius of 135μm for high conversion efficiencies typically required for applications such as narrow linewidth, multi-wavelength sources, or correlated photon pair generation [56, 57, 66]. In both cases the bus waveguides and the ring waveguide have the same cross section and fabrication process as previously described in Section 2.2 and 3 (see Fig. 3). The 4-port ring resonator is depicted in Fig. 10, and light is injected into the ring via vertical evanescence field coupling. The experimental set-up used to characterize the rings is shown in Fig. 11, and consists of 2 CW lasers, 2 polarizers, a power meter and a spectrometer. A Peltier cell is also used with the high Q ring for temperature control.

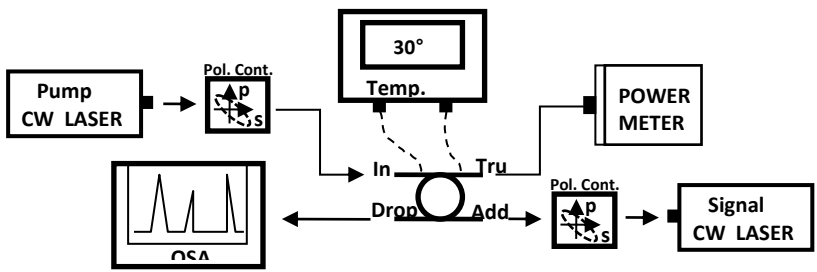

Fig. 11. Experimental set-up used to characterize the ring resonator and measure the converted idler from four wave mixing. 2 tunable fiber CW lasers are used, one at the input port and another at the drop port, whose polarizations and wavelengths are both set with inline fiber polarization controllers to match a ring resonance. The output spectrum and power are collected at the drop and through ports. A temperature controller is used to regulate the temperature of the device.

### 4.3.1 Dispersion

As detailed above, dispersion is a critical parameter in determining the efficiency of four-wave mixing. In ring resonators the dispersion can be directly extracted from the linear transmission through the ring. This was performed experimentally by using a wavelength tunable CW laser at the Input port and then recording the transmission at the drop port. The transmission spectral scan for the low Q ring can be seen in Fig. 12, from which a free spectral range of 575GHz and a Q factor of 65,000 were extracted (=200GHZ and 1,200,000 for the high Q).

As was derived in the beginning of Section 4, the propagation constants at resonance can be found to obey the relation: $\beta = m/R$, and thus are solely determined by the radius and an integer coefficient $m$. From vectorial finite element simulations the value of $m$ for a specific resonance frequency can be determined, and hence the integer value of all the experimentally determined resonances is obtained (as they are sequential). This provides a relation between the propagation constant $\beta$ and the angular frequency of the light $\omega$. By fitting a polynomial relation to this relation, as described by Eq. (3), the dispersion of the ring resonator is obtained. Fig. 13 presents the group velocity dispersion in the high Q ring (due to the smaller spectral range, a higher degree of accuracy was obtained here in comparison with the low Q) obtained by fitting a fourth order polynomial relation on the experimental data [68]. It is important to notice that the dispersion is extremely low for both the quasi-TE and quasi-TM modes of the structure, with zero dispersion crossings at $\lambda$ = 1560nm (1595nm) for the TM (TE) mode. At 1550nm we obtain an anomalous GVD of $\beta_2$= -3.1 ± 0.9 ps$^2$/km for the TM mode and -10 ± 0.9 ps$^2$/km for the TE mode. These low dispersion results were expected from previous considerations (see Section 3.1), and are ideal for four wave mixing applications.

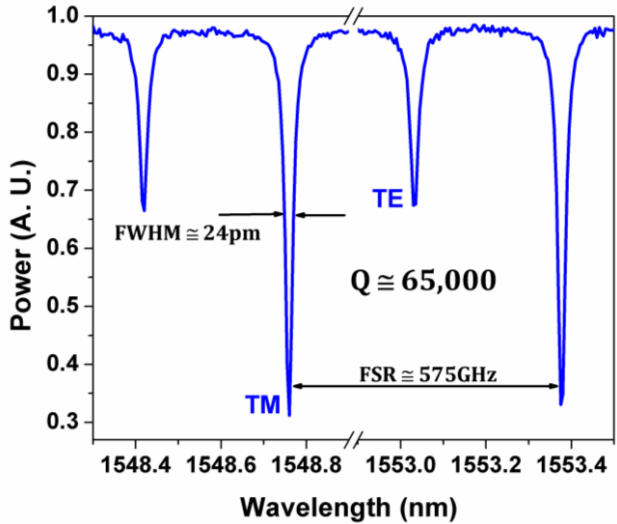

Fig. 12. Input-Drop response of the low Q (65,000) micro ring resonator.

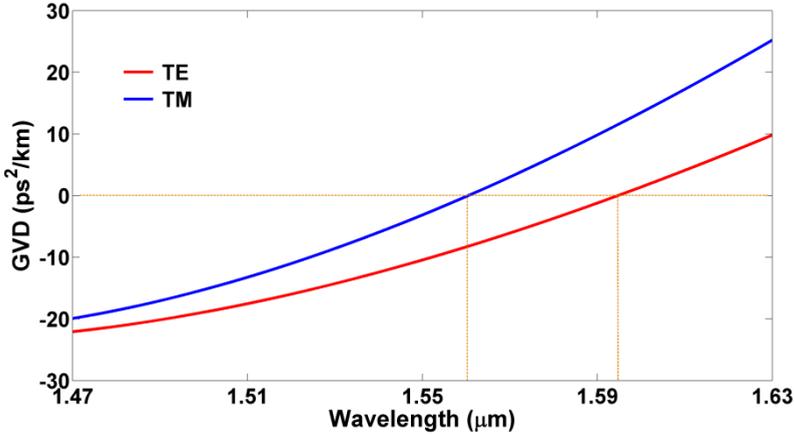

Fig. 13. Experimentally obtained dispersion (after fitting) of both the TE and TM fundamental modes. The zero-GVD points are found to be at 1594.7nm and 1560.5nm for TE and TM, respectively.

In addition, the dispersion data can be used to predict the bandwidth over which four-wave mixing can be observed. In resonators, the linear phase matching condition for the propagation constants is automatically satisfied as the resonator modes are related linearly by : $\beta = m/R$ [68, 69]. Rather, energy conservation becomes the new phase matching conditions, expressed as $\Delta\omega = 2\omega_2 - \omega_1 - \omega_{3r}$, where $\omega_1$ and $\omega_2$ are aligned to resonances by construction, but where the generated idler frequency $\omega_3 = 2\omega_2 - \omega_1$ is not necessarily aligned to its closest resonance $\omega_{3r}$ [68, 69]. We define the region where four-wave mixing is possible through the relation $|\Delta\omega| = |\omega_3 - \omega_{3r}| < \omega_{3r}/2Q$, such that the generated idler is within the bandwidth of the resonance. This condition for the high Q ring is presented in Fig. 14, where the region in black represents absence of phase matching, whereas the colored region represents possible four-wave mixing (the blue region implies the lowest phase mismatch, and red the highest). The curvature in the plots is a result of high order dispersive terms. It can be seen that the four wave mixing can be accomplished in the vicinity of the zero dispersion points up to 10THz (80nm) away from the pump. This extraordinary result comes from the low dispersion of the resonator, which permits appreciable phase matching over a broad bandwidth at low power. However, it is important to note that four-wave mixing can always be accomplished in an anomalous dispersion regime given a sufficiently high power [69].

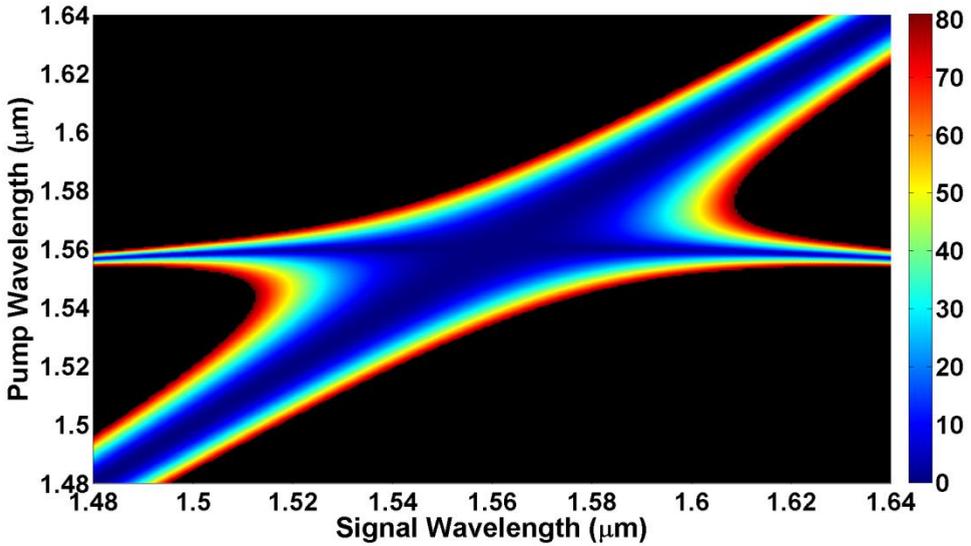

Fig. 14. Phase matching diagram associated to four-wave mixing in the high Q micro-ring resonator (interpolated). The regions in black are areas where four wave-mixing is not possible, whereas the coloured regions denote possible four-wave mixing with the colour indicating the degree of frequency mismatch (blue implies perfect phase matching; colour scale is $\Delta\omega$ in MHz).

### 4.3.2 Four-wave mixing in microring resonators

The 3 GHz bandwidth of the low Q resonator can be used for telecommunication applications such as signal regeneration and amplification [8, 9, 23]. We note that the device reported here was primarily designed for linear filter applications at 2.5 Gb/s, leaving room for further optimization for higher-bit-rate nonlinear applications. Four-wave mixing was obtained in this ring using only 5mW of input pump power at a resonance of 1553.38 nm, while the signal was tuned to an adjacent resonance at 1558.02nm with a power of 550 μW. Fig. 15 depicts the recorded output spectrum showing the generation of 2 idler frequencies: one of 930pW at 1548.74nm, and a second of 100pW at 1562.69nm. The latter idler is a result of formally exchanging the role of the pump and signal beams. The lower output idler power for the reverse process is a direct result of Eq. (16), where the conversion is shown to be proportional to the pump power squared. As is reported in [43], this is the first demonstration of four wave mixing in an integrated glass structure using CW light of such low power. This result is in part due to the relatively large $\gamma$ factor of 220 $W^{-1}km^{-1}$ (as compared to single mode fibers) and, more importantly, due to the low losses, resulting in a large intensity enhancement factor of $IE^4 \sim 1.4 \times 10^7$, which is orders of magnitude better than in semiconductors, where losses are typically the limiting factor [6, 27]. Recent results in SOI have also shown impressive, and slightly higher, conversion efficiencies using CW power levels. However, as can be seen in [6], saturation due to two-photon absorption generated free carriers limits the overall process, whereas in silica-doped glass it has been shown that no saturation effects occur for more than 25 $GW/cm^2$ of intensity, allowing for much higher conversion efficiencies with an increased pump power (note that the pump intensity in the ring is only ~0.015GW/cm² at resonance for 5mW of input power) [42].

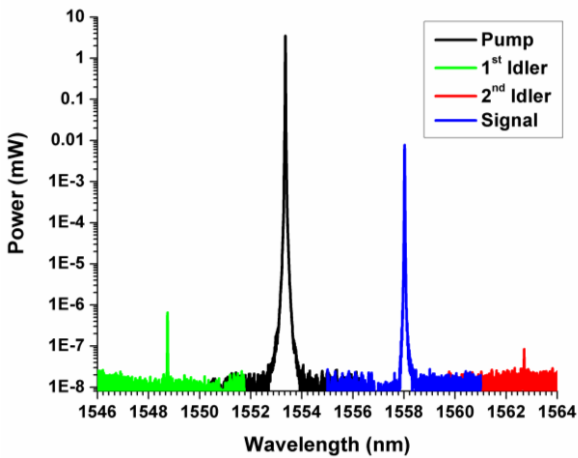

Fig. 15. Wavelength conversion in the low Q ring resonator.

The predicted frequency conversion, Eq. (16), was also verified experimentally. Firstly, for a fixed input pump power of 20mW the signal power was varied and the expected linear relationship between idler power and signal power was obtained, with a total conversion efficiency of 25 x 10$^{-6}$, as is shown in Fig. 16. Moreover, the reverse situation in which the pump power was varied for a fixed signal power also demonstrated the expected quadratic dependence, validating the approximations leading to Eq. (16). Lastly, by tuning the signal wavelength slightly off-resonance and measuring the conversion efficiency, it was experimentally shown, Fig. 17, that these results were in quasi-perfect phase matching, as predicted from Fig. 14.

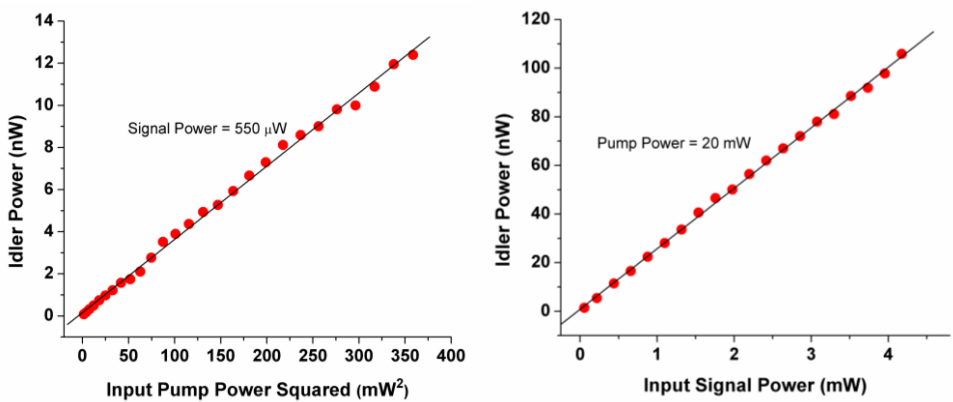

Fig. 16. Linear and quadratic relation of the idler power versus input signal power and input pump power, respectively.

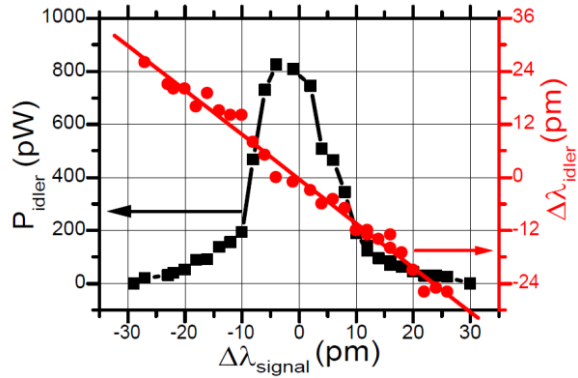

Fig. 17. Idler detuning curve, showing that dispersion is negligible in the system.

Experiments were also carried out in the high Q=1,200,000 ring resonator, which is for applications other than telecommunications, such as the realization of a narrow line source [70]. The advantage of this ring is the tremendous intensity enhancement factor $IE$ =302.8, which amounts to a conversion efficiency enhancement as high as 8.4 x $10^9$. Fig. 18 summarises the results of two different experiments where the pumps are placed to adjacent resonances, and when they are placed 6 free spectral ranges away from each other, respectively. In both cases the conversion efficiency was estimated to be -36 dB with only 8.8mW of input power. Moreover, a cascade of four-wave mixing processes can be seen whereby the pump and 1st idler mix to generate a 3rd idler (the numbers refer to Fig. 18).

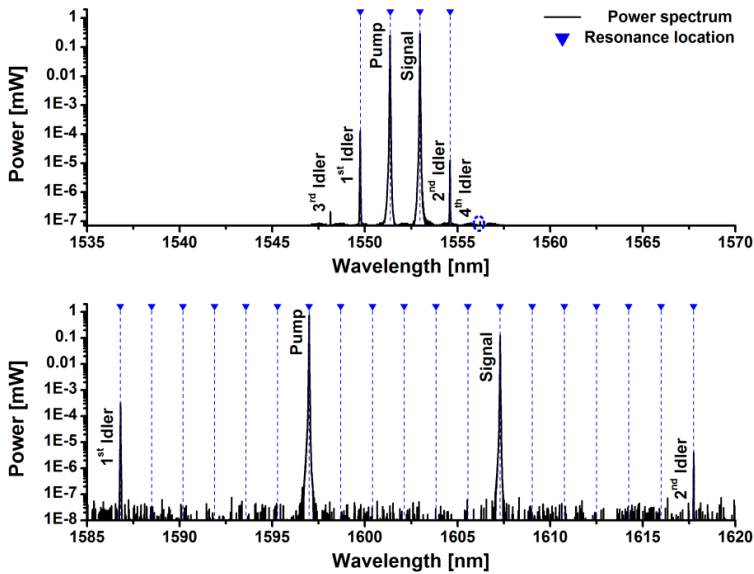

Fig. 18. Four wave mixing across several resonances in the high Q resonator. A conversion efficiency of -36db is obtained with only 8.8mW of external CW power.

## 5. Perspectives and conclusions

In this chapter we have presented a novel high-index doped-silica material platform for future integrated nonlinear optical applications. The platform acts as a compromise between the attractive linear properties of single mode fibers, namely low propagation losses and robust fabrication process, and those of semiconductors and other nonlinear glasses, and this by having a relatively large nonlinear parameter. Moreover, it outshines other high index glasses in its ability to have very low loss waveguides of 0.06 dB/cm without high temperature anneal, allowing for a complete CMOS compatible fabrication process. A small cross sectional area combined with a high index contrast also allows for tight bends down to 30 μm with negligible losses, permitting long spiral or resonant structures on chip. We have shown that although semiconductors possess a much larger nonlinearity $\gamma$, the low losses and robust fabrication allows for long and resonant structures with large and appreciable nonlinear effects that would otherwise not be possible in most semiconductors, or saturate with increasing input powers for others. In particular, we have presented and described measurement techniques to characterise the linear and third order nonlinearities, with specific applications to parametric four wave mixing.

Apart from the imminent applications in future photonic integrated circuits, these results may also pave way for a wide range of applications such as narrow linewidth, and/or multi-wavelength sources, on-chip generation of correlated photon pairs, as well as sources for ultra-low power optical hyper-parametric oscillators.